\documentclass[a4paper,12pt,preprint,nofootinbib,preprintnumbers,superscriptaddress]{revtex4-2}

\usepackage{amsmath}
\usepackage{graphicx,slashed,hyperref,color}
\usepackage{amssymb}
\usepackage[normalem]{ulem}
\usepackage[utf8]{inputenc}
\hypersetup{
   bookmarks=true,         
   unicode=true,          
   pdftoolbar=true,        
   pdfmenubar=true,        
   pdffitwindow=false,     
   pdfstartview={FitH},    
   pdftitle={My title},    
   pdfauthor={Author},     
   pdfsubject={Subject},   
   pdfcreator={Creator},   
   pdfproducer={Producer}, 
   pdfkeywords={keyword1} {key2} {key3}, 
   pdfnewwindow=true,      
   colorlinks=true,       
   linkcolor=blue,          
   citecolor=magenta,        
   filecolor=magenta,      
   urlcolor=cyan           
}

\def\lsim{\mathrel{\rlap{\lower4pt\hbox{\hskip1pt$\sim$}}
    \raise1pt\hbox{$<$}}}         
\def\gsim{\mathrel{\rlap{\lower4pt\hbox{\hskip1pt$\sim$}}
    \raise1pt\hbox{$>$}}}         

\numberwithin{equation}{section}

\begin{document}

{\flushright
\scriptsize
KIAS P22041\\ APCTP Pre2022 - 009\\ KYUSHU-HET-242\\ }

\title{Phenomenological implications on a hidden sector\\ 
from the Festina Lente bound}

\author{Kayoung Ban}
\affiliation{Department of Physics \& IPAP \& Lab for Dark Universe, Yonsei University,\\
Seoul 03722, Republic of Korea}

\author{Dhong Yeon Cheong}
\affiliation{Department of Physics \& IPAP \& Lab for Dark Universe, Yonsei University,\\
Seoul 03722, Republic of Korea}

\author{Hiroshi Okada}
\affiliation{Asia Pacific Center for Theoretical Physics (APCTP), Pohang 37673, Republic of Korea}
\affiliation{Department of Physics, Pohang University of Science and Technology,\\ 
Pohang 37673, Republic of Korea}

\author{Hajime Otsuka}
\affiliation{Department of Physics, Kyushu University, 744 Motooka, Nishi-ku,\\ 
Fukuoka 819-0395, Japan}

\author{Jong-Chul Park}
\email{jcpark@cnu.ac.kr}
\affiliation{Department of Physics and Institute of Quantum Systems (IQS), Chungnam National University, Daejeon 34134, Republic of Korea}

\author{Seong Chan Park}
\email{sc.park@yonsei.ac.kr}
\affiliation{Department of Physics \& IPAP \& Lab for Dark Universe, Yonsei University,\\
Seoul 03722, Republic of Korea}
\affiliation{Korea Institute for Advanced Study, Seoul 02455, Republic of Korea}

\begin{abstract}
We apply the Festina Lente (FL) bound on a hidden sector with $U(1)$ gauge symmetries.
Since the FL bound puts a lower bound on masses of particles charged under the $U(1)$ gauge symmetries, it is possible to constrain the hidden sector even with a tiny coupling to the Standard Model.
In particular, we focus on the phenomenological implications of the FL bound on milli-charged particles, which naturally arise when kinetic  mixing between the photon and the hidden photon  is allowed.
It turns out that the milli-charged particle with the mass $M\lesssim 5$\,meV is prohibited by the FL bound in the case of a single hidden $U(1)$, insensitively of the value of small kinetic mixing.
This bound is crucial when bosonic dark matter is taken in consideration in this framework: the fuzzy bosonic dark matter models requesting minuscule masses are ruled out by the FL bound if the longevity of dark matter is protected by the hidden gauge symmetry. 
\end{abstract}

\maketitle
\newpage

\section{Introduction}

The four-dimensional (4D) effective field theories (EFTs) admitting an ultra-violet completion to a consistent theory of quantum gravity will be characterized by the so-called swampland program~\cite{Vafa:2005ui,Brennan:2017rbf,Palti:2019pca}.
It states that 4D EFT satisfies several swampland conjectures arisen from consistent string compactifications as well as general principles.
One can constrain  particle physics models of inflation taking specific realization of swampland conjectures~\cite{Cheong:2018udx, Park:2018fuj}.

Among the proposed swampland conjectures, the weak gravity conjecture~\cite{Arkani-Hamed:2006emk} will be an established one, stating that any gauge force must be stronger than gravity.\footnote{See Ref.~\cite{Harlow:2022gzl} for a recent review about the weak gravity conjecture.} 
In a theory coupled to gravity with a $U(1)$ gauge symmetry, with a gauge coupling $g$, an electric version of the weak gravity conjecture puts on an upper bound on a particle 
in the theory with mass $m$ and charge $q$: $m < q g M_{\rm Pl}$. 
In the de Sitter (dS) space, this conjecture is valid when the size of the Reissner-Nordstr\"om dS black hole is much smaller than the dS radius.
On the other hand, when the black hole size is of the order of the dS radius, the masses of charged states ($m$) are lower bounded by the energy density of the vacuum, $V$~\cite{Montero:2019ekk,Montero:2021otb}: 
\begin{align}
 m^4 > 2q^2g^2 V
 \label{eq:FLsingle}
\end{align}
to avoid the superextremality.
This bound is called the Festina Lente (FL) bound. The FL bound puts highly nontrivial constraints on the phase structure of the Higgs field~\cite{Lee:2021cor} as the masses of particles are induced through the Higgs mechanism within the Standard Model (SM). 

Recalling that the current energy density of the Universe, $V \simeq (2.4 \times 10^{-3}~{\rm eV})^4$, the lower bound is close to the neutrino mass scale $m_\nu \sim 10^{-3}~{\rm eV}$.
However, we cannot directly apply the FL bound to the neutrinos, since they are uncharged under the electromagnetic $U(1)_{\rm em}$ of the SM gauge symmetries.
A gauged $U(1)_{B-L}$ model is an attractive scenario under which the neutrinos are charged particles, but only tiny $(B-L)$ gauge couplings are allowed to avoid the severe experimental constraints especially when neutrinos are Majorana fermions~\cite{Craig:2019fdy}.

In this paper, we seek phenomenological consequences of the FL bound for further extended cases with multiple $U(1)$ gauge symmetries. 
For concreteness, we first consider two unbroken Abelian gauge symmetries, including the $U(1)_{\rm em}$ and hidden $U(1)_H$ gauge fields with a small kinetic mixing.
Further extensions of our study must be straightforward. 
A kinetic mixing is a natural assumption since it is radiatively induced by bi-charged particles or string compactifications~\cite{Holdom:1985ag, Dienes:1996zr}.
From the phenomenological point of view, the existence of kinetic mixing leads to milli-charged particles (MCPs)~\cite{Holdom:1985ag,Feldman:2007wj,Huh:2007zw}.
So far, these particles are searched by several experiments and cosmological observations, but a wider region of parameter spaces is still allowed.
Our purpose is to put the FL bound on the MCPs and discuss phenomenological consequences.

This paper is organized as follows. 
In Sec. \ref{sec:FL}, we apply the FL bound on MCPs which are charged under unbroken $U(1)$ gauge symmetries and 
obtain the lower bound of MCP mass. 
In Sec. \ref{sec:pheno}, we present several upper bounds on the mixing parameter in terms of the MCP mass as 
summarized in Fig. \ref{fig:flexpup}. 
Sec. \ref{sec:con} is devoted to our conclusions and discussions. 
Another lower bound on the mass of fermionic DM is reviewed in Appendix \ref{sec:app}.

\section{Festina Lente Bound on Milli-Charged Particles}
\label{sec:FL}

A cosmological theory with multiple $U(1)$ gauge fields is described by 
the Lagrangian of the form 
\begin{align}
    e^{-1}{\cal L} = \frac{M_{\rm Pl}^2}{2}{\cal R}  
    - \Lambda_{\rm dS} - \frac{1}{4}f_{AB} F_{\mu\nu}^A F^{\mu\nu,B} - \mathcal{L}_{\rm MCP}\,, 
\end{align}
where $\Lambda_{\text{dS}} $ represents the dS vacuum energy density,  $F_{\mu \nu}^i$ is the gauge field-strength tensor for the $U(1)_i$ gauge field with the gauge coupling $g_i$, and $\mathcal{L}_{\text{MCP}}$ represents the milli-charged matter Lagrangian.
In principle, it can be either a fermion $\psi$ or a scalar particle $\phi$, each expressed with the Lagrangian
\begin{align}
    \mathcal{L}_{\rm MCP}^{\psi} &= \bar \psi (i \gamma^{\mu} D_{\mu} - M_{\psi} ) \psi\,,  \\
    \mathcal{L}_{\rm MCP}^{\phi} &=  |D_{\mu} \phi|^2 -M_\phi^2 |\phi|^2 - V(\phi)\,,
\end{align}
where the mass of $\psi (\phi)$ is denoted by $M_\psi (M_\phi)$.
The potential for $\phi$ is $V(\phi)$.

The FL bound for a charged particle with the mass $M(\equiv M_\psi$ or $M_\phi)$ is extended to the covariant expressions with respect to multiple fields~\cite{Montero:2021otb}:
\begin{align}
    M^4 > 2 q_A^\prime (f^{-1})^{AB} q_B^\prime \Lambda_{\rm dS} = 6 q_A^\prime (f^{-1})^{AB} q_B^\prime (M_{\rm Pl}^2 H^2)\,,
    \label{eq:FL}
\end{align}
where $q_{i}^\prime \equiv q_i g_i$ ($i=A,B$), $\Lambda_{\text{dS}}[=V$ in Eq.~(\ref{eq:FLsingle})] $\equiv 3(M_{\text{Pl}} H)^2\simeq (2.4 \times 10^{-3}{\rm eV})^{4}$, $M_{\rm Pl}$ is the reduced Planck mass, and $H$ is the Hubble parameter.
One can check that the above general form includes the single field case, Eq.~(\ref{eq:FLsingle}), replacing $f^{-1}$ with 1.

Taking a model that has two Abelian gauge symmetries, $U(1)_{\rm em}$ in the SM and a hidden $U(1)_H$ where $U(1)_H$ is an unbroken symmetry,\footnote{See Ref.~\cite{Chun:2010ve} for the case of a broken $U(1)_H$.} the valid kinetic terms of the electromagnetic and hidden gauge fields,
$A'_\mu$ and $A'_{H\mu}$, mix each other, and are given as follows~\cite{Huh:2007zw,Park:2012xq}: 
\begin{align}
{\cal L}&=
-\frac14 F'_{\mu\nu}F'^{\mu\nu}
-\frac14 F'_{H\mu\nu}F'^{\mu\nu}_H
-\frac\chi2 F'_{H\mu\nu}F'^{\mu\nu}\,, 
\label{eq:fs}
\end{align}
where $F'_{\mu\nu}$ and $F'_{H\mu\nu}$ are gauge field strengths of  $U(1)_{\rm em}$ and  $U(1)_H$, respectively.
Then, we write the canonical basis $-{\cal L}=\frac14 F_{\mu\nu}F^{\mu\nu} +\frac14 F_{H\mu\nu}F_H^{\mu\nu}$ by transforming $A'_\mu$ and $A'_{H\mu}$ as follows: 
\begin{align}
A'_\mu=\frac{A_\mu}{\sqrt{1-\chi^2}},\quad
A'_{H\mu}=A_{H\mu} -\frac{\chi}{\sqrt{1-\chi^2}} A_\mu\,.
\end{align}
Next, we consider a $U(1)_\text{em}$ neutral fermion $\psi$ that has a hidden charge $q_H^\psi$ with a mass $M_\psi$.
Then, the gauge interaction term in $A'_\mu$ and $A'_{H\mu}$ basis, $\mathcal{L} \supset \bar \psi\gamma_\mu ( g_H q_H^\psi A'^\mu_H)\psi$, is rewritten in terms of canonical basis as follows: 
\begin{align}
\mathcal{L} \supset \bar \psi\gamma_\mu (g_{\rm mix} q_H^\psi A^\mu + g_H q_H^\psi A^\mu_H)\psi\,, \,\, \quad g_{\rm mix}= - \frac{g_H \chi}{\sqrt{1-\chi^2}}\,,
\end{align}
where  $g_H$ represents the $U(1)_H$ gauge coupling.
Here, we assume that $\psi$ is a singlet under the $SU(2)_L$ gauge group.
Note that the same analogy applies for a scalar $\phi$ as well, with our notations denoting $g_H q_H^{\phi} $ for the bosonic case.

Under the setup, the FL bound for $\psi,\ \phi$ is evaluated as follows:
\begin{align}
M_{\psi, \phi}^4
\gtrsim
\frac{6(M_\text{Pl} H)^2}{1-\chi^2} (g_H q_H^{\psi, \phi})^2\,,
\end{align}
where we have used 
$ f = \begin{pmatrix}
    1 & \chi \\
    \chi & 1
    \end{pmatrix}$ 
in Eq.~(\ref{eq:FL}).
Note that the dependence on the mixing parameter enters in the denominator.
In case of $\chi \ll 1$,\footnote{Realizing a very small kinetic mixing may be prone to theoretical issues, and explicit model building may be required. 
See e.g. Ref.~\cite{Gherghetta:2019coi} and references therein.} the FL bound is simplified to the form
\begin{align}
\begin{split}
M_{\psi, \phi}^4
\gtrsim
{6(M_{\text{Pl}} H)^2} (g_H q_H^{\psi, \phi})^2  \rightarrow M_{\psi, \phi} \gtrsim 5 \times \left(\frac{g_H q_H^{\psi,\phi}}{\sqrt{4\pi}}\right)^{1/2}~\text{meV}\,.
\end{split}
\end{align}
The result now solely depends on the gauge charge of the hidden $U(1)_H$.
It suggests that the maximum value of the right handed side is $\sim 5$ meV when a perturbative limit is saturating with $g_H q_H^{\psi, \phi}\approx \sqrt{4\pi}$.
The scale is intriguingly the same as the one of neutrino masses.

One can generalize the results for multiple hidden gauge symmetries, $[U(1)_H]^N$: the FL bound is generalized to be
\begin{align}
    M_{\psi, \phi}^4 \gtrsim 6(M_{\text{Pl}} H)^2 \sum_{j=1}^N (g_{H, j} q_{H, j}^{\psi, \phi})^2\,.
\end{align}
If the gauge coupling constants and the charges are about the same size, $g_{H,j} q_{H,j} \sim g_H q_H$, and they saturate the perturbative bound $\sim \sqrt{4\pi}$, 
the lower FL bound increases by the factor of $N^{1/4}$ compared to the case of the single $U(1)_H$ symmetry. 

If the MCP is the only particle charged under $U(1)_H$, it will be subject to the weak gravity conjecture, hence there will be an upper bound on the particle's mass at $g_H q_H^{\psi, \phi} M_\text{Pl}$. Therefore, one finds the window for a consistent MCP:
\begin{align}
    \left(\frac{6}{1 - \chi^2}\right)^{1/4} \sqrt{ g_H q_H^{\psi, \phi} M_{\text{Pl}} H}  \lesssim M_{\psi, \phi} \lesssim g_H q_H^{\psi, \phi} M_\text{Pl}\,.
    \label{eq:bound}
\end{align}
This theoretical bound is one of our most important results for a MCP.

\section{Phenomenological Implications and Experimental Constraints} \label{sec:pheno}

Since there have been extensive searches for MCP, we now want to discuss phenomenological implications of the bound (\ref{eq:bound})   with existing experimental constraints.

\subsection{Experimental and observational bounds on MCPs}

Here, we present a compilation of the constraints on MCPs. 
\begin{itemize}

\item MCP as Dark Matter \newline
If the MCP is the lightest among all the particles charged under the hidden $U(1)_H$, its stability is automatically guaranteed.
Therefore, it is a natural candidate of dark matter (DM) even though it may not be the dominant component of DM~\cite{Dasgupta:2021ies}. 
The fermionic DM is subject to the Tremaine-Gunn bound~\cite{Tremaine:1979we} since the phase space density is confined under the degenerate Fermi gas due to the Pauli exclusion principle. (See Appendix for details.)
A lower bound on bosonic DM mass comes from the fact that its wave length should not be greater than the size of the halos of dwarf galaxies ~\cite{Schive:2015kza}.
Recently, a more stringent constraint on bosonic DM has been reported based on the modeling of the Milky Way (MW) satellite galaxy population and the abundance of observed MW satellites from the Dark Energy Survey and Pan-STARRS1~\cite{DES:2020fxi}.
\begin{align}
M_{\rm DM} \gsim \begin{cases}
100~\text{eV}, &\text{Tremaine-Gunn (Fermion)} \\
1.2 \times 10^{-22}~\text{eV}, &\text{dwarf galaxy size (Boson)}\\
2.9 \times 10^{-21}~\text{eV}, &\text{MW satellite galaxy population (Boson)}\,. 
\end{cases}
\end{align}
We note that the FL bound is automatically satisfied for a fermionic DM when the Tremaine-Gunn bound is applied, but a large parameter space for a bosonic DM (fuzzy DM) is excluded by the FL bound. 

\item Red Giants and White Dwarfs~\cite{Davidson:1991si} \newline
These bounds correspond to the impact of the MCPs on the stellar evolution of the relevant astrophysical objects.
Due to the mixing of $U(1)_H$ with the $U(1)_\text{em}$, MCP pairs emitted by plasmon decay provide an additional stellar energy loss channel, changing helium-related processes in the reflection grating spectrometer and further accelerating the cooling of white dwarfs.
This bound, given that the production of these MCP pairs will not be restricted by a small mass, extends all the way down the mass window.

\item SN1987A~\cite{Chang:2018rso} \newline
This bound, also obtained from the modified stellar evolution due to MCPs, is derived from the change in the neutrino emission of the star based on the `Raffelt criterion'.
Improved considerations at the higher mass end greatly extend previous considerations.

\item BBN ($N_{\text{eff}}^{\text{BNN}}$) and CMB ($N_{\text{eff}}^{\text{CMB}}$)~\cite{Vogel:2013raa} \newline
The Big Bang nucleosynthesis (BBN) and cosmic microwave background (CMB) bounds correspond to the constraints in the extra relativistic degree of freedom or $\Delta N_{\text{eff}}$ measured in each era.
As these particles, however infinitesimal, carry small SM electric charges due to the kinetic mixing, MCPs can interact with the SM thermal plasma, further producing more relativistic degree of freedoms.
The most relevant production mechanism from the thermal relativistic plasma are 
$$e^{+} e^{-} \rightarrow \bar{\psi}{\psi}\,, \quad e\psi \rightarrow e\psi$$
and, they provide the $\epsilon^2$ suppressed results in $\Delta N_{\text{eff}} \approx 0.69\times 10^{17}\times \epsilon^2$.
From the currently obtained bounds for the CMB and the BBN era are $N_{\text{eff}}^{\text{CMB}} = 2.99_{-0.33}^{+0.34}$ \cite{Planck:2018vyg} and $N_{\text{eff}}^{\text{BNN}}=2.85^{+0.28}_{-0.28}$ \cite{Cyburt:2015mya}, we can set the bound on the mixing angle $\epsilon\lesssim 10^{-9}$. 

\item SLAC (Beam Dump)~\cite{Prinz:1998ua}  \newline
An experiment uniquely suited to the production and detection of such MCPs has been carried out at SLAC.
This experiment is sensitive to the infrequent excitation and ionization of matter expected from the passage of such a particle.
This analysis rules out a region of mass and charge with 95\% confidence upper limit.

\item WMAP (Wilkinson Microwave Anisotropy Probe)~\cite{Dubovsky:2003yn} \newline
Using the CMB data from WMAP and assuming the standard BBN value for the baryon abundance, $\Omega_{b} h^2_0 = 0.0214 \pm 0.0020$ \cite{Kirkman:2003uv}, the constraint on the MCP abundance is $\Omega_{\text{MCP}} h^2_0 < 0.007$ (95\% C.L.) if MCPs are coupled to baryons at the recombination epoch.

\item DM Relic Abundance ($\Omega_{\rm DM}h^2 > 0.1$)~\cite{Davidson:1991si} \newline
The DM constraints will apply to the paraphoton (dark photon $U(1)_H$) which could not have cooled out of the halo within the age of the Universe, $\epsilon^2 < 7 \times 10^{-11} \frac{M_\psi}{n_e}$.
Here $n_e$ is the free electron number density and if all the electrons are free, $n_e \sim 1$.

\item Other Bounds include Vacuum Biref., Accelerator cavities, Ortho-positronium, Lamb Shift, and Accelerators~\cite{Essig:2013lka,Bhandari:2021dsh}.
\end{itemize}

%
\begin{figure}[t]
\centering
\includegraphics[width=16cm]{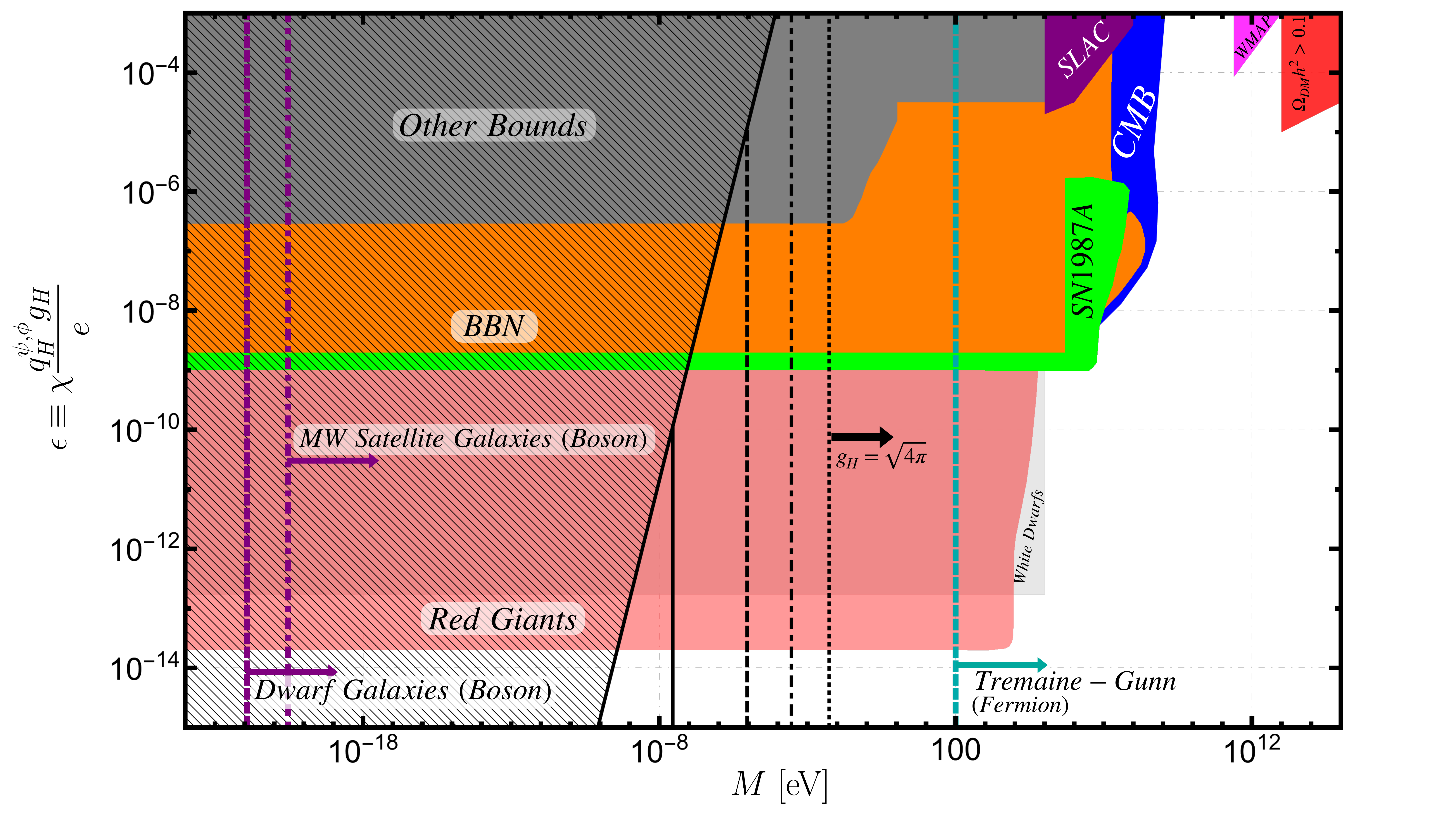}
\caption{
Constraints on the electric charge of MCP $\epsilon$ in terms of its mass $M$.
The FL bounds for $g_H q_H^{\psi, \phi} =(10^{-10},~ 10^{-5},~ 10^{-2},~ \sqrt{4\pi})$ are respectively shown as solid, dashed, dot-dashed, and dotted vertical black lines.
The shaded diagonal region corresponds to the kinetic mixing $\chi > 1$, which is ruled out of our theory space.
The thick dashed cyan line is the Tremaine-Gunn bound~\cite{Tremaine:1979we} which applies for fermions, while the dashed and dot-dashed purple lines correspond to the lower bounds on bosonic DM mass from the size of the halos of dwarf galaxies~\cite{Schive:2015kza} and the MW satellite galaxy population~\cite{DES:2020fxi}, respectively.
In addition, we depict a compilation of bounds from red giants, white dwarfs, SN1987A, BBN, CMB, SLAC, WMAP, and DM relic abundance ($\Omega_{\rm DM} h^2 > 0.1$) in different colors~\cite{Goodsell:2009xc,Davidson:1991si,Chang:2018rso,Vogel:2013raa,Essig:2013lka,Bhandari:2021dsh,Dubovsky:2003yn,Prinz:1998ua}. 
}
\label{fig:flexpup}
\end{figure}

Taking all these constraints discussed above, in Fig.~\ref{fig:flexpup}, we collect all the bounds on the electric charge of MCP, $\epsilon \equiv \chi \frac{q_H^{\psi, \phi} g_H}{e}$, in terms of its mass $M$~\cite{Abel:2008ai}, along with the constraints given by the FL bound.
The vertical black solid, dashed, dot-dashed, and dotted lines respectively represent the FL bounds for the cases of $g_H q_H^{\psi, \phi} =(10^{-10},~ 10^{-5},~ 10^{-2},~ \sqrt{4\pi})$, which correspond to $M \gtrsim (2.9\times10^{-8},~ 9.0\times 10^{-6},~ 2.9\times10^{-4},~ 5.4\times10^{-3}) $ eV.
The figure suggests that MCP $\psi$ is severely restricted in terms of the electrical charge $\epsilon$ and the most stringent bound comes from RGs, $\epsilon \lesssim 10^{-13}$ for $M \lesssim 10~{\rm keV}$.
Note that essentially the same bounds are obtained for a scalar MCP $\phi$.
As we are considering a kinetic mixing between the $A_\mu^{\prime}$ and $A_{H\mu}^{\prime}$, parameter regions exhibiting $\chi>1$ are forbidden. 
Also, for $g_H q_H^{\psi, \phi} \ll 1$ the restrictions from the FL bound weaken depending on its precise value.

\section{Conclusions and Discussions}
\label{sec:con}

In this paper, we  studied phenomenological implications of the Festina-Lente (FL) bound on the hidden sector gauge symmetries and milli-charged particles (MCP) associated with them.
In the case of a $[U(1)_H]^N$ with $N\geq 1$ and universal $\mathcal{O}(1)$ couplings, we found that the FL bound is set as
\begin{align}
M \gsim 5 N^{1/4} ~{\rm meV},
\end{align}
insensitively of the value of small kinetic mixing.
This bound is already interesting, as the lightest charged particle of unbroken $U(1)_H$ is a DM candidate.
We saw that extremely light bosonic (fuzzy) dark matter $m_\phi \sim 10^{-21}$ eV, if protected by a hidden gauge symmetry, is ruled out by the FL bound for any reasonable values of charge and coupling.
On the other hand, fermionic dark matter is allowed satisfying the Tremaine-Gunn (TG) bound around 100 eV.
We also compared the FL bound with all the available experiments and astrophysical observations for direct and indirect searches of MCP.
We found that the FL bound, which is insensitive to the value of small kinetic mixing, is the most stringent upper bound in the region of $M\lesssim 10^4$ eV when $\epsilon \lsim 10^{-14}$.

\acknowledgments

This work is supported by an appointment to the JRG Program at the APCTP through the Science and Technology Promotion Fund and Lottery Fund of the Korean Government and by the Korean Local Governments - Gyeongsangbuk-do Province and Pohang City (Hiroshi O.). 
This work is supported by JSPS KAKENHI Grant Numbers JP20K14477 (Hajime O.). 
The work is supported by the National Research Foundation of Korea (NRF) [NRF-2021R1A4A2001897 (JCP, SCP),  NRF-2019R1C1C1005073 (JCP), NRF-2019R1A2C1089334 (SCP)].

\appendix
\section{Lower Bound on the Mass of Fermionic Dark Matter}
\label{sec:app}

In this section, we review the lower bound of milli-charged fermionic DM masses, based on~\cite{Tremaine:1979we, Boyarsky:2008ju, Davoudiasl:2020uig}.
The standard nomenclature regarding the mass bound on light leptons is the Tremaine-Gunn (TG) bound~\cite{Tremaine:1979we}.
This bound can be generalized to fermions: 
due to the Pauli exclusion principle, the stacking of fermions in a certain phase space region is limited.
The requirement of the DM phase space density to be confined under the degenerate Fermi gas leads to a lower mass bound.
For simplicity and as a crude estimate, if one assumes a spherically symmetric DM dominated object with mass $M$ within a region $R$, one can find the following condition by requiring the maximal Fermi velocity does not exceed the escape velocity:
\begin{align}
\hbar \left(\frac{9 \pi M}{2 g m_\text{DEG}^4 R^3} \right)^{1/3} \leq \sqrt{\frac{2 G_N M}{R}}  ~\rightarrow~ m_\text{DEG}^4 \geq \frac{9 \pi \hbar^3}{4\sqrt{2} g M^{1/2} R^{3/2} G_N^{3/2}} 
\end{align}
which, by inserting specific values from observations~\cite{Boyarsky:2008ju}, provides a limit on the $M_{\psi}$ as
\begin{align}
M_\psi \gsim m_{\text{DEG}} \sim 100 ~\text{eV}\,.
\end{align}

\bibliography{ref}{}
\bibliographystyle{JHEP} 

\end{document}